\documentclass{article}

\usepackage[english]{babel}
\usepackage[a4paper,top=2cm,bottom=2cm,left=3cm,right=3cm,marginparwidth=1.75cm]{geometry}

\usepackage{natbib}
\usepackage{amsmath}
\usepackage{amssymb}
\usepackage{amsthm}
\usepackage{mathtools}\mathtoolsset{showonlyrefs=true}
\usepackage{graphicx}
\usepackage[colorlinks=true, allcolors=blue]{hyperref}

\usepackage{caption}
\usepackage{subcaption}

\usepackage{dsfont}

\newcounter{scenario}[section]

\providecommand{\keywords}[1]{\textbf{\textit{Keywords---}} #1}

\newcommand{\RR}{\mathbb{R}} % R
\newcommand{\TT}[1]{\mathcal{T}_{#1}} % Domain definition space
\newcommand{\HH}{\mathcal{H}} % Product of L^p space
\newcommand{\pointt}{\mathbf{t}} % Indexed of multivariate curves
\newcommand\restr[2]{{ %
  \left.\kern-\nulldelimiterspace  %
  #1  %
  \vphantom{\big|}  %
  \right|_{#2}  %
}}
\title{On the estimation of the number of components in multivariate functional principal component analysis}
\author{%
Steven Golovkine\thanks{MACSI, Department of Mathematics and Statistics, University of Limerick, Ireland \href{mailto:steven.golovkine@ul.ie}{steven.golovkine@ul.ie}}
\and
Edward Gunning\thanks{MACSI, Department of Mathematics and Statistics, University of Limerick, Ireland \href{mailto:edward.gunning@ul.ie}{edward.gunning@ul.ie}}
\and
Andrew J. Simpkin\thanks{School of Mathematical and Statistical Sciences, University of Galway, Ireland \href{mailto:andrew.simpkin@nuigalway.ie}{andrew.simpkin@nuigalway.ie}}
\and
Norma Bargary\thanks{MACSI, Department of Mathematics and Statistics, University of Limerick, Ireland \href{mailto:norma.bargary@ul.ie}{norma.bargary@ul.ie}}
}
\date{\today}

\begin{document}
\maketitle

\begin{abstract}
\cite{happMultivariateFunctionalPrincipal2018} developed a methodology for principal components analysis of multivariate functional data observed on different dimensional domains. Their approach relies on an estimation of univariate functional principal components for each univariate functional feature. In this paper, we present extensive simulations to investigate choosing the number of principal components to retain. We show empirically that the conventional approach of using a percentage of variance explained threshold for each univariate functional feature may be unreliable when aiming to explain an overall percentage of variance in the multivariate functional data, and thus we advise practitioners to exercise caution.
\end{abstract}


% MAIN --------

\section{Introduction} % (fold)
\label{sec:introduction}

\cite{happMultivariateFunctionalPrincipal2018} develop innovative theory and methodology for the dimension reduction of multivariate functional data on possibly different dimensional domains (e.g., curves and images). Their work extends existing methods that were limited to either univariate functional data or multivariate functional data on a common one-dimensional domain. Recent research has shown a growing presence of data defined on different dimensional domains in diverse fields such as biomechanics, e.g., \cite{warmenhovenBivariateFunctionalPrincipal2019} and neuroscience, e.g., \cite{songSparseMultivariateFunctional2022}, therefore there are significant practical applications for their methods. We aim to provide guidance on the estimation of the number of principal components utilising the methodology proposed in \cite{happMultivariateFunctionalPrincipal2018}. This is discussed briefly in \citet[Online Supplement, Section 2.3]{happMultivariateFunctionalPrincipal2018}, where the authors note the influence of the univariate decomposition on the estimation of the multivariate principal components. To achieve this, we conduct an extensive simulation study and subsequently propose practical guidelines for practitioners to adeptly choose the appropriate number of components to retain for multivariate functional datasets. For ease of presentation, we use the same notation as in \cite{happMultivariateFunctionalPrincipal2018}. Code to reproduce the simulation study and data analysis in this discussion is available at \url{https://github.com/FAST-ULxNUIG/variance_mfpca}.

% section introduction (end)

\section{Model} % (fold)
\label{sec:model}

\cite{happMultivariateFunctionalPrincipal2018} proposed an extension of functional principal components analysis (FPCA, \cite{ramsayFunctionalDataAnalysis2005}) to multivariate functional data defined on different dimensional domains, named multivariate functional principal components analysis (MFPCA). The data, referred to as multivariate functional data, consist of independent trajectories of a vector-valued zero-mean stochastic process $X = (X^{(1)}, \ldots, X^{(p)}),~p \geq 1$. For each $1 \leq j \leq p$, let $\mathcal{T}_j \in \RR^d$ with $d \geq 1$. Each feature $X^{(j)}: \mathcal{T}_j \longrightarrow \RR$ is assumed to be in $L^{2}(\mathcal{T}_j)$. We define the matrix of covariance functions $C(\cdot, \cdot) = \mathbb{E}(X(\cdot) \otimes X(\cdot))$ with elements
\begin{equation}\label{eq:cov}
C_{ij}(s_i, t_j) = \mathbb{E}(X^{(i)}(s_i)X^{(j)}(t_j)), \quad s_i \in \mathcal{T}_i, t_j \in \mathcal{T}_j.    
\end{equation}
MFPCA consists of decomposing the covariance structure of the multivariate functional data into a set of orthogonal basis functions, named the (multivariate) principal components. Let $x_1, \dots, x_N$ be $N$ realisations of the process $X$. We briefly summarise the estimation procedure of the principal components given the sample $x_1, \dots, x_N$ with full details provided in \cite{happMultivariateFunctionalPrincipal2018}, Section 3. For all $n = 1, \dots, N$, the observation $x_n$ is a vector of $p$ functions. We denote by $x_n^{(j)}$ the $j$th entry of the vector $x_n$, referred to as the $j$th functional feature of the $n$th observation. The first step is to perform a univariate FPCA for each individual feature $X^{(j)}$ using $x_1^{(j)}, \dots, x_N^{(j)}$. For a component $X^{(j)}$, the eigenvalues and eigenfunctions are computed as a matrix analysis of the estimated covariance $C_{jj}$ from~\eqref{eq:cov}. We estimate $M_j$ univariate functional principal components for each feature $j$. This results in a set of eigenfunctions $(\phi^{(j)}_{1}, \ldots, \phi^{(j)}_{M_j})$ associated with a set of eigenvalues $(\lambda^{(j)}_1, \ldots, \lambda^{(j)}_{M_j})$ for each feature $j$. The univariate scores for a realisation $x^{(j)}_n$ of $X^{(j)}$ are then given by $\xi^{(j)}_{n, m} = \langle  x^{(j)}_n, \phi^{(j)}_m \rangle_2,~m = 1, \ldots, M_j$ where $\langle \cdot, \cdot\rangle_2$ is the usual inner-product in $L^2(\mathcal{T}_j)$. The total number of components that have been estimated over all $p$ features is thus $M_+ = \sum_{j = 1}^p M_j$. We also define $M_{-} = \min_{j = 1, \dots, p} M_j$ to be the minimum number of univariate components estimated across all univariate features $j$. By concatenating the scores obtain for the $p$ features, we obtain a matrix $\Xi \in \RR^{N \times M_+}$ where each row $(\xi^{(1)}_{n, 1}, \ldots, \xi^{(1)}_{n, M_1}, \ldots, \xi^{(p)}_{n, 1}, \ldots, \xi^{(p)}_{n, M_p})$ contains the estimated scores for a single realisation. An estimation of the covariance of the matrix $\Xi$ is given by $\mathbf{Z} = (N - 1)^{-1}\Xi^\top\Xi$. An eigenanalysis of the matrix $\mathbf{Z}$ is performed resulting in eigenvalues $\nu_m$ and eigenvectors $\mathbf{c}_m$. Finally, the multivariate eigenfunctions are estimated as a linear combination of the univariate eigenfunctions using
\begin{equation}
    \psi^{(j)}_m(t_j) = \sum_{n = 1}^{M_j}[\mathbf{c}_m]^{(j)}_n \phi^{(j)}_{n}(t_j), \quad t_j \in \mathcal{T}_j, \quad m = 1, \dots, M_+,
\end{equation}
where $[\mathbf{c}_m]^{(j)}_n$ denotes the $n$th entry of the $j$th block of the vector $\mathbf{c}_m$. The multivariate eigenvalues are the same as the eigenvalues $\nu_m$ of the matrix $\mathbf{Z}$. The multivariate scores are estimated as
\begin{equation}
    \rho_{n,m} = \Xi_{n, \cdot}\mathbf{c}_m, \quad n = 1, \ldots, N, \quad m = 1, \ldots, M_+,
\end{equation}
where $\Xi_{n, \cdot}$ is the $n$th row of the matrix $\Xi$. In the context of the paper, our focus lies in investigating how the selection of the parameters $M_j$ impacts the estimation of the eigenvalues $\nu_m$ and the estimation of the eigenfunctions $\psi_m$.

Using this methodology, the maximum number of multivariate principal components that can be estimated is $M_+$. Let $\{\nu_m\}_{1 \leq m \leq M_+}$ be the set of true eigenvalues and $\{\widehat{\nu}_m\}_{1 \leq m \leq M_+}$ be the set of estimated eigenvalues. We use the relative errors $\text{Err}(\widehat{\nu}_m)  = (\nu_m - \widehat{\nu}_m)^2 / \nu^2_m,~m = 1, \ldots, M_+$ to assess the accuracy of the estimates. In \cite{happMultivariateFunctionalPrincipal2018}, the authors also propose to estimate the number of multivariate components using the percentage of variance explained. For that, they first select $M_j$ univariate components that explain $\alpha\%$ of the variance for each univariate feature \cite[Chapter 8.2]{ramsayFunctionalDataAnalysis2005} and they claim that this number of components is enough to estimate the number of multivariate components that explain $\alpha\%$ of the variance in the multivariate functional data \cite[Section 3.2]{happMultivariateFunctionalPrincipal2018}. Using a simulation study, we show that selecting $M_j$ univariate components that explain $\alpha\%$ of the variance for each individual feature can lead to an under-estimation of the number of multivariate principal components required to explain $\alpha\%$ of the variance over all features. The percentage of variance explained by the $m$th component and the cumulative percentage of variance explained by the first $m$ components are defined as
\begin{equation}\label{eq:pve}
     \text{PVE}_m = 100 \times \nu_m \times \left(\sum_{l = 1}^{M_+} \nu_l\right)^{-1} \quad\text{and}\quad \text{PVE}_{1:m} = \sum_{l = 1}^m \text{PVE}_l, \quad m = 1, \dots, M_+.
\end{equation}
If we fix the percentage of variance explained to be $\alpha\%$, the number of components needed to explain $\alpha\%$ of the variance is given by
\begin{equation}\label{eq:npc}
     \text{NPC}_{\alpha} = \sum_{m = 1}^{M_{+}} \mathbf{1}\left\{\text{PVE}_{1:m} < \alpha\right\} + 1.
\end{equation}

% section model (end)

\section{Simulation} % (fold)
\label{sec:simulation}

We perform a simulation study based on the first setting presented in the simulation study conducted in \cite{happMultivariateFunctionalPrincipal2018}. In this scenario, the data-generating process is based on a truncated version of the Karhunen-Loève decomposition. First, we generate a large orthonormal basis $\{\psi_m\}_{1 \leq k \leq M}$ from $L^{2}(\mathcal{T})$ on an interval $\mathcal{T} = [0, T] \subset \RR$. We fix $T_1 = 0$ and $T_{p + 1} = T$ and we generate $p - 1$ cut points $T_2, \dots, T_p$ uniformly in $\TT{}$ such that $0 = T_1 < \cdots < T_p < T_{p+1} = T$. Let $s_1, \dots, s_p \in \{-1, 1\}$ be coefficients that randomly flip the eigenfunctions with probability $0.5$, generated according to a Bernoulli distribution. The univariate components of the eigenfunctions are then defined as
\begin{equation}\label{eq:simulation_uni_component}
    \psi_m^{(j)}(t_j) = s_j \restr{\psi_m}{[T_j, T_{j + 1}]}\left(\frac{t_j - T_j}{T_{j + 1} - T_j}\right), \quad m = 1, \dots, M, \quad j = 1, \dots, p.
\end{equation}
The notation $\restr{\psi_m}{[T_j, T_{j + 1}]}$ is the restriction of the function $\psi_m$ to the set $[T_j, T_{j + 1}]$. The set of multivariate functions $\{\psi_m\}_{1 \leq m \leq M}$ is an orthonormal system in $\HH \coloneqq L^2({\mathcal{T}_{1}}) \times \dots \times L^2({\mathcal{T}_{p}})$ with $\mathcal{T}_{j} = [0, 1]$. Each curve is then simulated using the truncated multivariate Karhunen-Loève expansion,
\begin{equation}
    x_n(\pointt) = \sum_{m = 1}^M \rho_{i, m} \psi_m(\pointt), \quad \pointt \in \TT{},\quad n = 1, \dots, N,
\end{equation}
where the scores $\rho_{n, m}$ are sampled as Gaussian random variables with mean $0$ and variance $\nu_m$. The eigenvalues $\nu_m$ are defined with an exponential decrease, $\nu_m = \exp(-(m + 1)/2)$. We simulate $N = 25, 50$ and $100$ observations for each replication of the simulation. Similarly, each component is sampled on a regular grid of $S = 25, 50$ and $100$ sampling points. We use $p = 5$ features and we set $M = 50$. This estimation procedure consists of densely observed multivariate functional data defined on different one-dimensional domains. The parameters are chosen to reflect the sample size and observation points typically found in real-world datasets. The estimation is done using the \textsf{R} package \texttt{MFPCA} (\cite{happ-kurzObjectOrientedSoftwareFunctional2020}). For each univariate feature $j$, we estimate $M_j$ principal components. Then, following the multivariate components estimation procedure, we can estimate $M_+ = \sum_{j = 1}^p M_j$ multivariate components. The simulations are replicated $500$ times.

\begin{figure}
     \centering
    \includegraphics[width=0.94\textwidth]{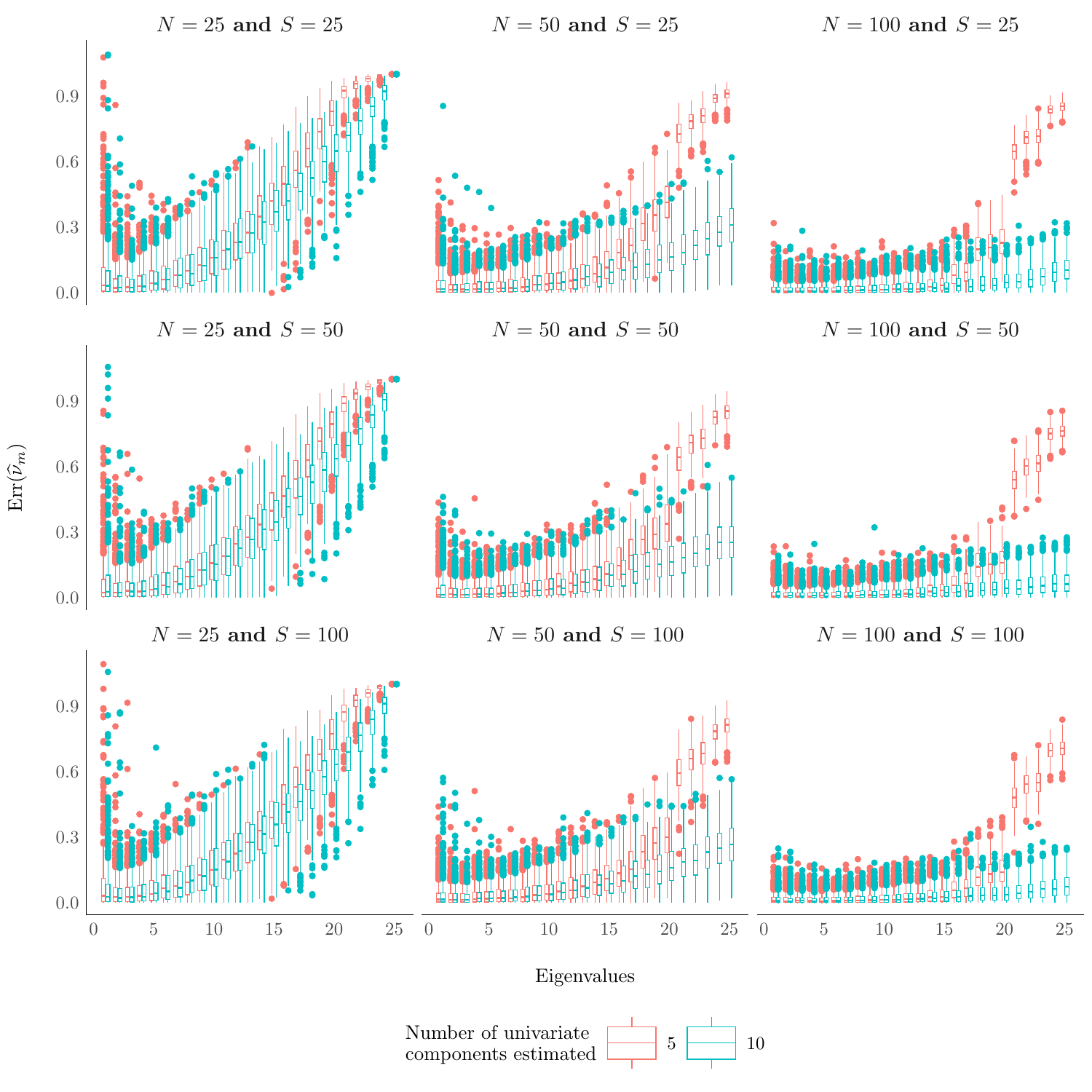}
    \caption{Boxplots of the estimation errors of the eigenvalues. We estimated $M_j = 5$ (red boxplots) and $M_j = 10$ (blue boxplots) univariate functional components for $j = 1, \dots, p$. The number of multivariate principal components that are estimated is $25$. $N$ is the number of observations, $S$ is the number of sampling points per curve. We run $500$ simulations.}
    \label{fig:ncomp}
\end{figure}

To illustrate the effect of $M_j$ on the estimation of the eigenvalues $\nu_m$, Figure \ref{fig:ncomp} displays a comparison of the errors for the first $25$ estimated eigenvalues $\widehat{\nu}_m$ when using $M_j = 5$ and $M_j = 10$ for all $j = 1, \ldots, p$. The accuracy of the estimation of the multivariate eigenvalue $\widehat{\nu}_m$ declines with $m$ in all scenarios. However, the decrease in accuracy is faster when $M_j = 5$ than when $M_j = 10$. We observe in particular a notable drop in accuracy for the estimated eigenvalues $\widehat{\nu}_m, m = 21, \dots, 25$ when $M_j = 5$ in most scenarios, while this drop does not appear when $M_j = 10$. The dependence of the choice of $M_j$ (the number of univariate functional principal components retained) on the multivariate eigenvalue estimates $\widehat{\nu}_m$ is clear. Increasing $M_j$ improves the accuracy of the multivariate eigenvalue estimates, and for a given $m$, there is a required minimum number of univariate principal components that should be retained to accurately estimate the multivariate eigenvalues. However, in our simulation, the errors in the estimation of the first five multivariate eigenvalues are similar when $M_j = 5$ and $M_j = 10$. Hence, it is also not useful to use too large a value for $M_j$ as this will only increase the computational effort required. We suggest to estimate at most $M_{-}$ multivariate components using $M_j$ univariate components; otherwise, the univariate components may not contain enough information to effectively recover their corresponding multivariate counterparts.

In a second setting, for each univariate feature $j$, we fix a percentage of variance to be explained $\alpha\%$ for each univariate feature and choose $M_j$ principal components accordingly. We then estimate $M_+$ multivariate components and replicate the simulations $500$ times. Table~\ref{fig:npc_estim} presents the estimation of the number of multivariate components $\widehat{\text{NPC}}_{\alpha}$ retained across $500$ simulation scenarios. The quantity $\text{NPC}_{\alpha}$ represents the number of multivariate components that would be needed to explain at least $\alpha\%$ of the variance ($50\%, 70\%, 90\%, 95\%$ and $99\%$), considering an exponential decay of the eigenvalues as defined in~\eqref{eq:npc}. Note that, we can compute the true number of multivariate components as we know the true eigenvalues. Each entry in Table~\ref{fig:npc_estim} indicates the number of times each number of multivariate components has been selected over the $500$ simulations. The number of components appears to be consistently underestimated for various combinations of the number of observations $N$, number of sampling points $S$, and desired percentage of variance explained $\alpha\%$. Therefore, this simulation scenario shows that using a percentage of variance explained of level $\alpha\%$ to choose $M_j$ is not sufficient to estimate the number of multivariate functional principal components that explain $\alpha\%$ of the variance in the multivariate functional data. These findings may hold considerable significance for practitioners as the percentage of variance explained by each eigencomponent is a popular method to determine the number of principal components retained (see, e.g., \cite{jamesIntroductionStatisticalLearning2021} for scalar data and \cite{horvathInferenceFunctionalData2012a} for functional data).
\begin{table}
\begin{subtable}[h]{0.3\textwidth}
\centering
\begin{tabular}{rr|rr}
    & & \multicolumn{2}{c}{$\widehat{\text{NPC}}_{\alpha}$} \\
    $N$ & $S$ & 1 & 2 \\
    \hline
    25 & 25 & \textbf{301} & 199 \\
    25 & 50 & \textbf{276} & 224 \\
    25 & 100 & \textbf{270} & 230 \\
    50 & 25 & 235 & \textbf{265} \\
    50 & 50 & 208 & \textbf{292} \\
    50 & 100 & 254 & \textbf{246} \\
    100 & 25 & 158 & \textbf{342} \\
    100 & 50 & 165 & \textbf{335} \\
    100 & 100 & 178 & \textbf{322} \\
\end{tabular}
\caption{$\alpha = 50\%$ ($\text{NPC}_{\alpha} = 2$)}
\end{subtable}
\hfill
\begin{subtable}[h]{0.3\textwidth}
\centering
\begin{tabular}{rr|rrr}
    & & \multicolumn{3}{c}{$\widehat{\text{NPC}}_{\alpha}$} \\
    $N$ & $S$ & 1 & 2 & 3\\
    \hline
    25 & 25 & 1 & \textbf{464} & 35\\
    25 & 50 & 1 & \textbf{456} & 43\\
    25 & 100 & 1 & \textbf{461} & 38\\
    50 & 25 & 0 & \textbf{459} & 41\\
    50 & 50 & 0 & \textbf{461} & 39\\
    50 & 100 & 1 & \textbf{450} & 49\\
    100 & 25 & 0 & \textbf{467} & 33\\
    100 & 50 & 0 & \textbf{469} & 31\\
    100 & 100 & 0 & \textbf{471} & 29\\
\end{tabular}
\caption{$\alpha = 70\%$ ($\text{NPC}_{\alpha} = 3$)}
\end{subtable}
\hfill
\begin{subtable}[h]{0.3\textwidth}
\centering
\begin{tabular}{rr|rrr}
    & & \multicolumn{3}{c}{$\widehat{\text{NPC}}_{\alpha}$} \\
    $N$ & $S$ & 3 & 4 & 5 \\
    \hline
    25 & 25 & 15 & \textbf{400} & 85\\
    25 & 50 & 18 & \textbf{375} & 107\\
    25 & 100 & 12 & \textbf{379} & 109\\
    50 & 25 & 0 & \textbf{322} & 178\\
    50 & 50 & 0 & \textbf{271} & 229\\
    50 & 100 & 0 & \textbf{268} & 232\\
    100 & 25 & 0 & 212 & \textbf{288}\\
    100 & 50 & 0 & 157 & \textbf{343}\\
    100 & 100 & 0 & 136 & \textbf{364}\\
\end{tabular}
\caption{$\alpha = 90\%$ ($\text{NPC}_{\alpha} = 5$)}
\end{subtable}
\\
\begin{subtable}[h]{0.45\textwidth}
\centering
\begin{tabular}{rr|rrrr}
    & & \multicolumn{4}{c}{$\widehat{\text{NPC}}_{\alpha}$} \\
    $N$ & $S$ & 4 & 5 & 6 & 7 \\
    \hline
    25 & 25 & 6 & \textbf{379} & 115 & 0\\
    25 & 50 & 5 & \textbf{376} & 118 & 1\\
    25 & 100 & 1 & \textbf{357} & 142 & 0\\
    50 & 25 & 0 & \textbf{288} & 212 & 0\\
    50 & 50 & 0 & 232 & \textbf{267} & 1\\
    50 & 100 & 0 & 210 & \textbf{289} & 1\\
    100 & 25 & 0 & 172 & \textbf{328} & 0\\
    100 & 50 & 0 & 110 & \textbf{390} & 0\\
    100 & 100 & 0 & 84 & \textbf{416} & 0\\
\end{tabular}
\caption{$\alpha = 95\%$ ($\text{NPC}_{\alpha} = 6$)}
\end{subtable}
\hfill
\begin{subtable}[h]{0.45\textwidth}
\centering
\begin{tabular}{rr|rrrr}
    & & \multicolumn{4}{c}{$\widehat{\text{NPC}}_{\alpha}$} \\
    $N$ & $S$ & 7 & 8 & 9 & 10\\
    \hline
    25 & 25 & 10 & \textbf{365} & 125 & 0\\
    25 & 50 & 2 & \textbf{362} & 136 & 0\\
    25 & 100 & 2 & \textbf{338} & 160 & 0\\
    50 & 25 & 0 & 117 & \textbf{383} & 0\\
    50 & 50 & 0 & 86 & \textbf{413} & 1\\
    50 & 100 & 0 & 52 & \textbf{448} & 0\\
    100 & 25 & 0 & 8 & \textbf{492} & 0\\
    100 & 50 & 0 & 0 & \textbf{499} & 1\\
    100 & 100 & 0 & 2 & \textbf{497} & 1\\
\end{tabular}
\caption{$\alpha = 99\%$ ($\text{NPC}_{\alpha} = 10$)}
\end{subtable}
\caption{Estimation of the number of components to explain $\alpha\%$ of the variance over $500$ simulations. The true number of components that explain $\alpha\%$ of the variance is given in parenthesis. $N$ is the number of observations, $S$ is the number of sampling points per curve.}
\label{fig:npc_estim}
\end{table}

% section simulation (end)

\section{Application: Canadian weather dataset} % (fold)
\label{sec:application_canadian_weather_dataset}

\begin{figure}
     \centering
     \begin{subfigure}[b]{0.49\textwidth}
         \centering
         \includegraphics[width=1\textwidth]{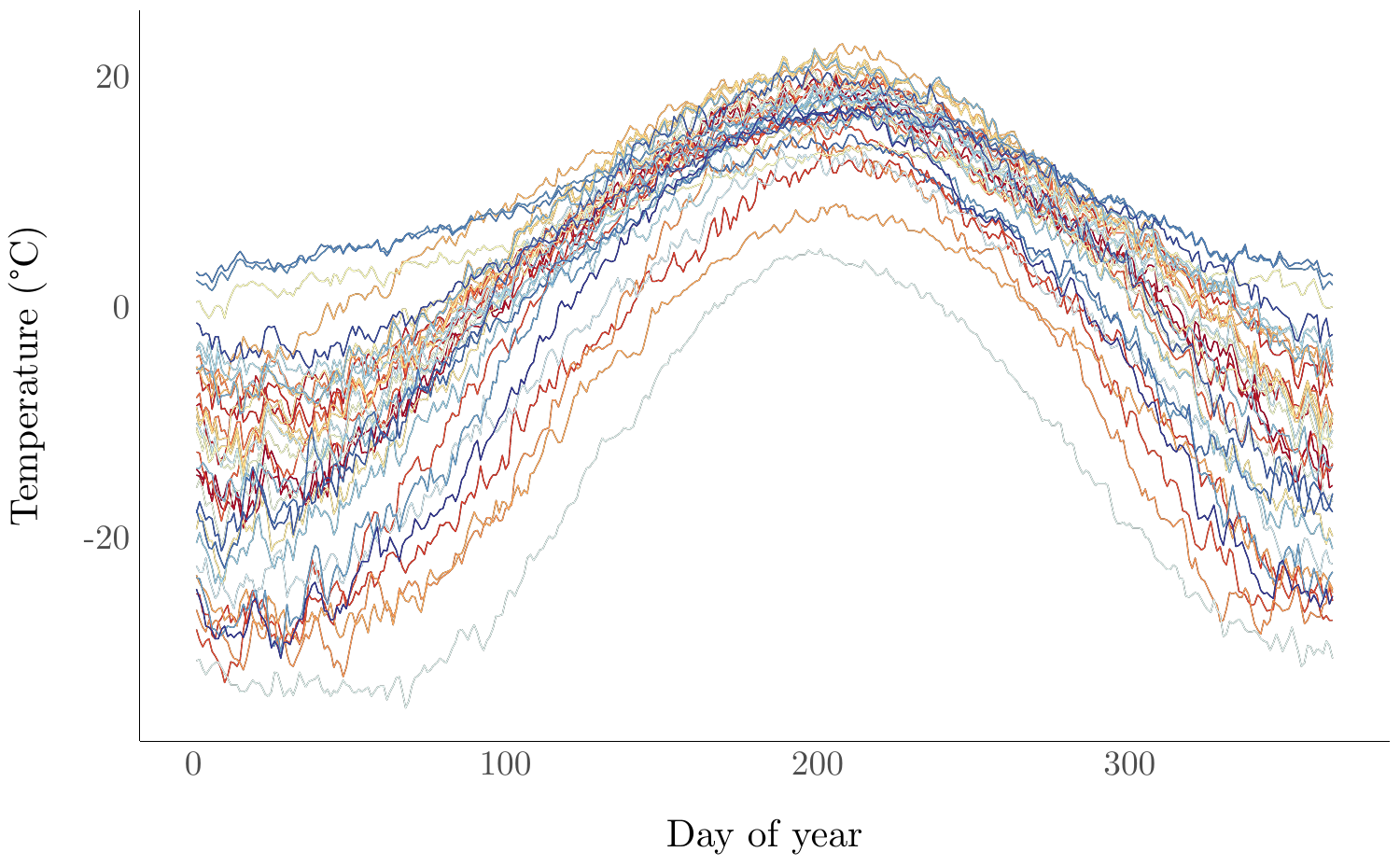}
         \caption{Temperature (first feature)}
         \label{fig:temperature}
     \end{subfigure}
     \hfill
     \begin{subfigure}[b]{0.49\textwidth}
         \centering
         \includegraphics[width=1\textwidth]{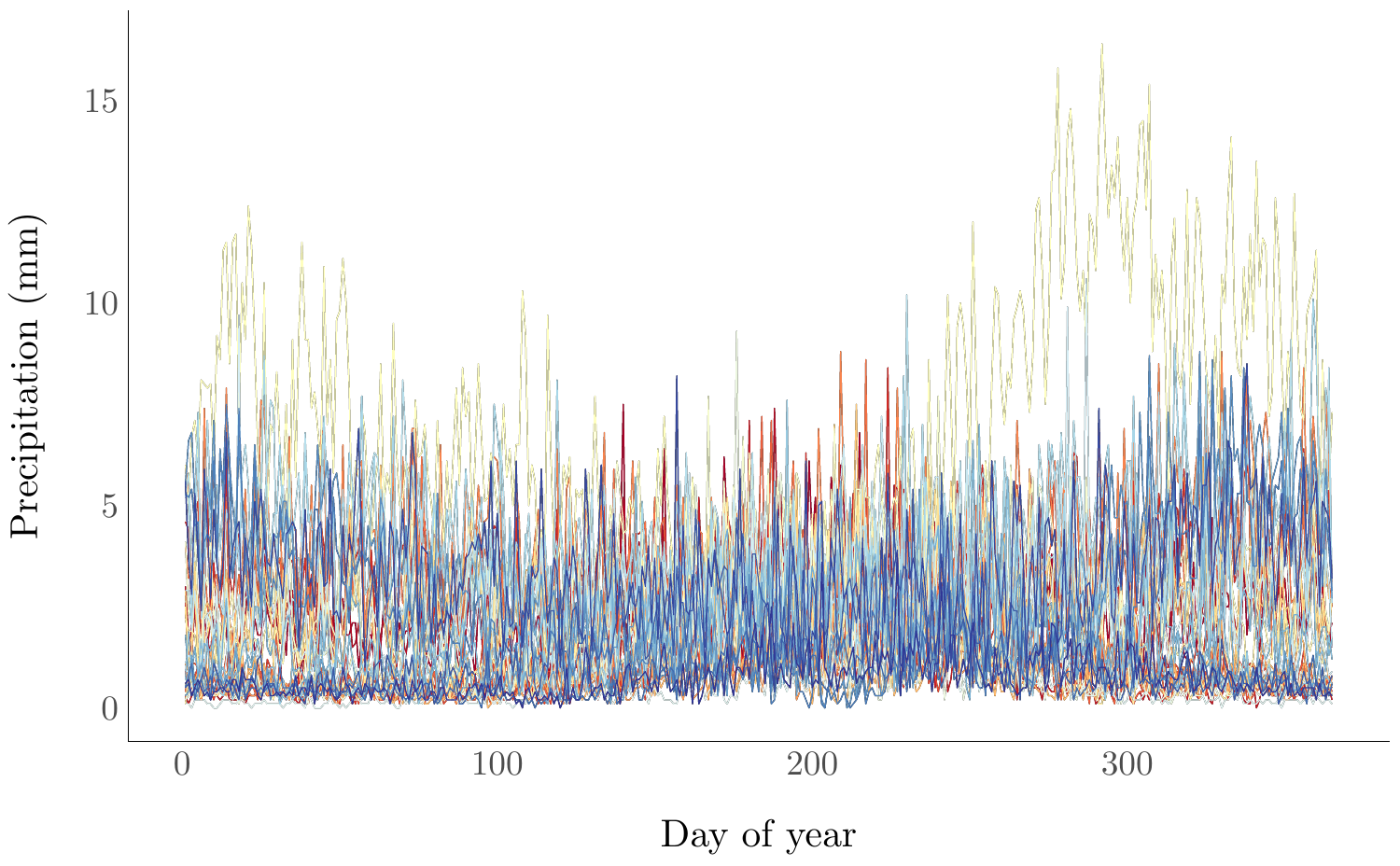}
         \caption{Precipitation (second feature)}
         \label{fig:precipitation}
     \end{subfigure}
     \caption{The daily temperature and precipitation in $35$ Canadian weather stations. Each curve represents one weather station.}
     \label{fig:weather}
\end{figure}

To illustrate our simulation results, we apply the same idea on a real dataset, the Canadian weather dataset \citep{ramsayFunctionalDataAnalysis2005}, available in the \textsf{R} package \texttt{fda} \citep{ramsayFdaFunctionalData2023}. The dataset contains daily measurements of temperature (in Celsius) and precipitation (in millimeters) for $35$ Canadian weather stations, averaged over the years 1960 to 1994. The data are presented in Figure~\ref{fig:weather}. This is an example of multivariate functional data with $p = 2$ defined on one dimensional domains, the temperature being the first feature $x^{(1)}$ and the precipitation being the second feature $x^{(2)}$. We aim to estimate $M$ multivariate principal components of the data using different numbers of univariate principal components $M_j$ and compare the results. We define two scenarios, one where $M = M_{+}$ and one where $M = M_{-}$.

We first expand the data in a B-spline basis with $10$ basis functions. In the first scenario, for each feature, we estimate two univariate principal components ($M_1 = 2$ and $M_2 = 2$). In the second scenario, for each feature, we estimate four univariate principal components ($M_1 = 4$ and $M_2 = 4$). In both scenarios, we then estimate $M = 4$ multivariate principal components. So, for the first scenario, $M = M_{+} = 4$ and for the second scenario, $M = M_{-} = 4$. Based on the simulation and Figure \ref{fig:ncomp}, we expect the first two multivariate principal components to be similar and the other two to be different.

\begin{figure}
     \centering
     \begin{subfigure}[b]{0.49\textwidth}
         \centering
         \includegraphics[width=1\textwidth]{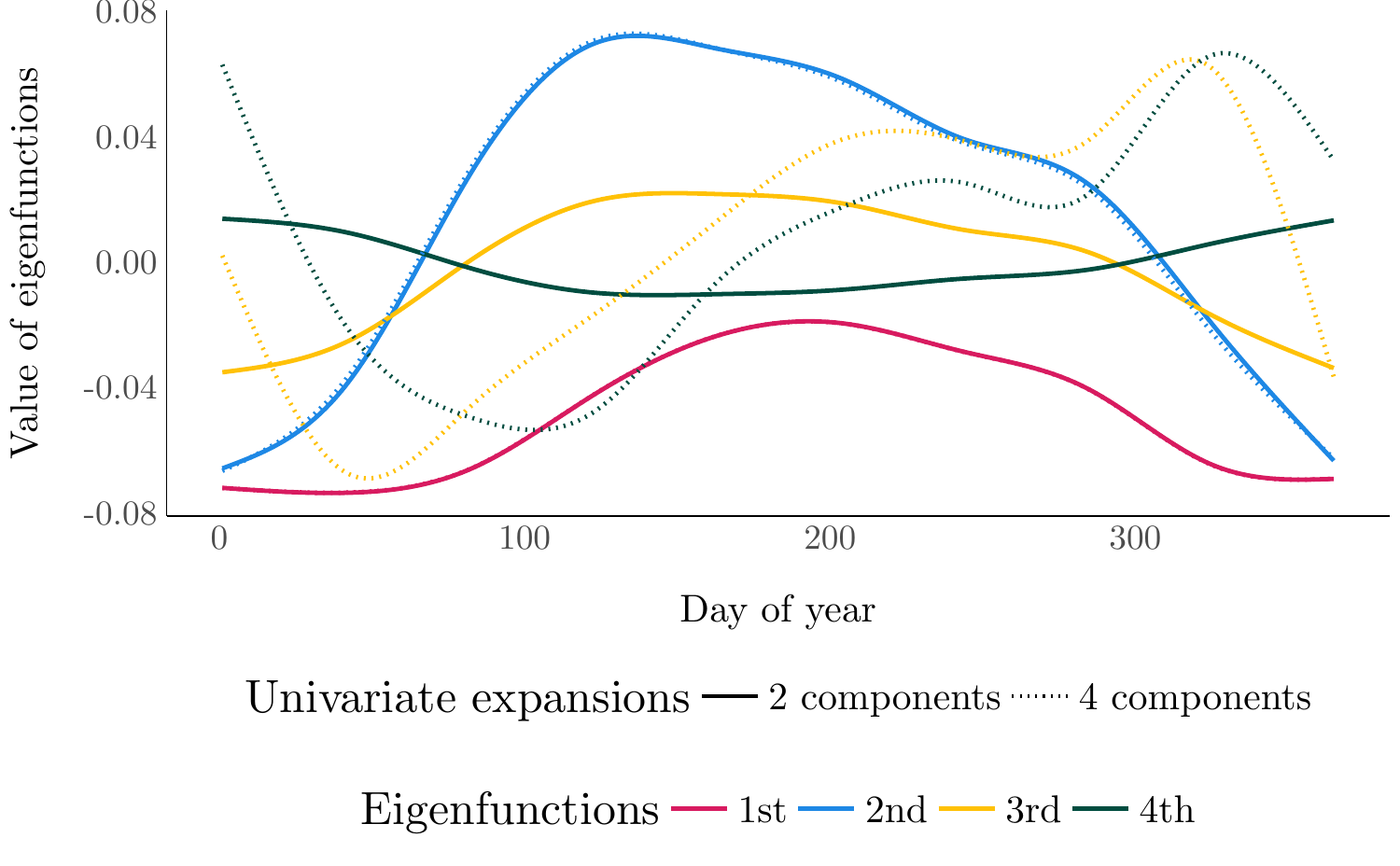}
         \caption{Temperature (first component)}
         \label{fig:temperature}
     \end{subfigure}
     \hfill
     \begin{subfigure}[b]{0.49\textwidth}
         \centering
         \includegraphics[width=1\textwidth]{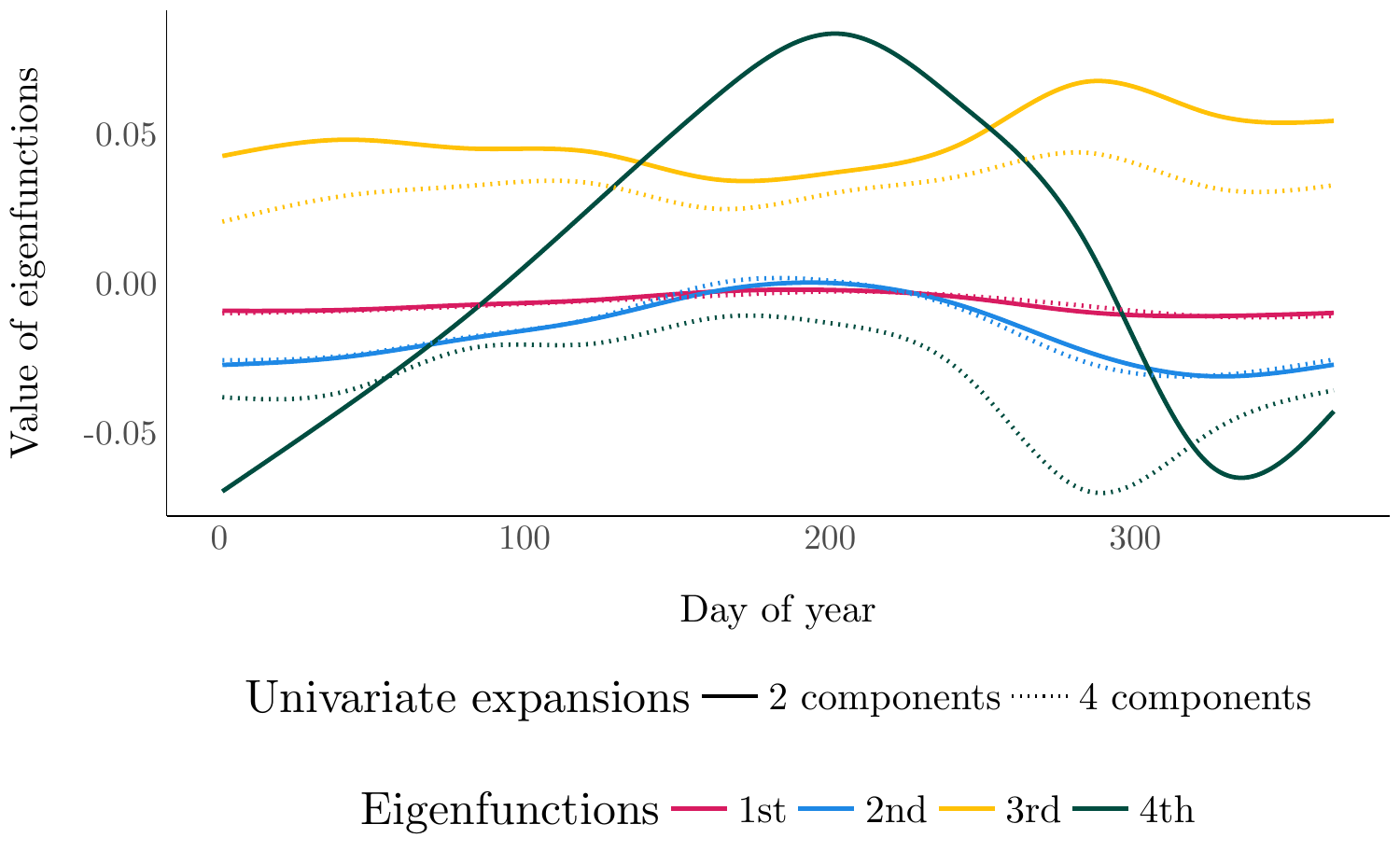}
         \caption{Precipitation (second component)}
         \label{fig:precipitation}
     \end{subfigure}
     \caption{Estimation of the first four eigenfunctions of the Canadian weather dataset using two and four univariate components.}
     \label{fig:eigenfunctions_weather}
\end{figure}

Table \ref{tab:eigenvalues} presents the estimation of the eigenvalues for the Canadian weather dataset for both scenarios. We notice that the values are similar for the first two eigenvalues, but quite different for the other two. 
\begin{table}
\centering
\begin{tabular}{c c | c c c c}
 & & \multicolumn{4}{c}{Eigenvalues} \\
Scenario & Univariate expansions & 1st & 2nd & 3rd & 4th \\
\hline
1 & $2$ components & $15845$ & $1675$ & $308$ & $45$ \\
2 & $4$ components & $15850$ & $1679$ & $438$ & $213$ \\
\hline
\end{tabular}
\caption{Estimation of the first four eigenvalues of the Canadian weather dataset using two and four univariate components.}
\label{tab:eigenvalues}
\end{table}
Figure \ref{fig:eigenfunctions_weather} presents the estimation of the eigenfunctions for the Canadian weather dataset for both scenarios. As with the eigenvalues, we notice that the first two (multivariate) eigenfunctions are approximately the same, but the other two are not. The first two principal components can be interpreted similarly in both scenarios. The first component is negative for both features, indicating that weather stations with positive scores have lower temperatures and less precipitation than average. While the first component for precipitation is relatively flat, the temperature component exhibits more variation at the beginning and end of the year. The second component contrasts winter and summer: stations with positive scores have higher temperatures and more precipitation in summer, and lower temperatures with less precipitation in winter compared to the average station. The third and fourth components differ between the scenarios. For univariate expansions with two components, the third multivariate component for temperature contrasts winter and summer, while with four univariate components, it contrasts spring and autumn. The third multivariate component for precipitation varies in magnitude depending on the number of univariate components. The fourth multivariate component for temperature is roughly flat when two univariate components are used but shows more variability with four components. For precipitation, the fourth component contrasts winter and summer when using two univariate components, but becomes flatter with a negative bump in autumn when four components are considered. These results highlight that the interpretation of the multivariate components depends on the number of univariate components used for the univariate decomposition. However, it would be preferable that the interpretation remains consistent and independent of the univariate decomposition. While we do not know the true eigenfunctions, based on simulation results, the estimated multivariate eigenfunctions from the second scenario should be closer to the truth than the estimation from the first scenario. In the first scenario, the univariate decompositions did not capture enough information to accurately estimate the third and fourth multivariate principal components. We reiterate the suggestion to estimate at most $M_{-}$ multivariate principal components (which is $M_{-} = 2$ for the first scenario).

% section application_canadian_weather_dataset (end)

\section{Conclusion} % (fold)
\label{sec:conclusion}

\cite{happMultivariateFunctionalPrincipal2018} present a general methodology to estimate principal components for a set of multivariate functional data defined on, possibly, different dimensional domains. Their approach, based on the decomposition of the covariance of each univariate feature, allows easy estimation of the components.

We have conducted a simulation study and an example on a real dataset, and the obtained results highlight two important findings. Firstly, although utilising only a few univariate components may yield a substantial number of multivariate components, their accuracy is notably limited. Secondly, relying on the percentage of variance explained as a criterion for selecting the number of univariate components may result in an underestimation of the number of multivariate components. We therefore advise practitioners to exercise caution when determining the number of estimated components required in their analysis. We suggest to estimate at most $M_{-}$ multivariate components if for each univariate feature, $M_j$ univariate components have been estimated. Additionally, we strongly recommend conducting simulations that closely resemble the characteristics of the actual data to select the appropriate number of components based on the percentage of variance explained criterion.

% section conclusion (end)
% -------------

% ACKNOWLEDGMENT -------
\section*{Acknowledgment}

S. Golovkine, A. J. Simpkin and N. Bargary are partially supported by Science Foundation Ireland under Grant No. 19/FFP/7002 and co-funded under the European Regional Development Fund. E. Gunning is supported in part Science Foundation Ireland (Grant No. 18/CRT/6049) and co-funded under the European Regional Development Fund. The authors also wish to acknowledge the Irish Centre for High-End Computing (ICHEC) for the provision of computational facilities and support.

\bibliographystyle{plainnat}
\bibliography{./biblio}

\begin{thebibliography}{8}
\providecommand{\natexlab}[1]{#1}
\providecommand{\url}[1]{\texttt{#1}}
\expandafter\ifx\csname urlstyle\endcsname\relax
  \providecommand{\doi}[1]{doi: #1}\else
  \providecommand{\doi}{doi: \begingroup \urlstyle{rm}\Url}\fi

\bibitem[Happ and Greven(2018)]{happMultivariateFunctionalPrincipal2018}
Clara Happ and Sonja Greven.
\newblock Multivariate {{Functional Principal Component Analysis}} for {{Data
  Observed}} on {{Different}} ({{Dimensional}}) {{Domains}}.
\newblock \emph{Journal of the American Statistical Association}, 113\penalty0
  (522):\penalty0 649--659, April 2018.
\newblock ISSN 0162-1459.
\newblock \doi{10.1080/01621459.2016.1273115}.

\bibitem[{Happ-Kurz}(2020)]{happ-kurzObjectOrientedSoftwareFunctional2020}
Clara {Happ-Kurz}.
\newblock Object-{{Oriented Software}} for {{Functional Data}}.
\newblock \emph{Journal of Statistical Software}, 93:\penalty0 1--38, April
  2020.
\newblock ISSN 1548-7660.
\newblock \doi{10.18637/jss.v093.i05}.

\bibitem[Horv{\'a}th and Kokoszka(2012)]{horvathInferenceFunctionalData2012a}
Lajos Horv{\'a}th and Piotr Kokoszka.
\newblock \emph{Inference for {{Functional Data}} with {{Applications}}},
  volume 200 of \emph{Springer {{Series}} in {{Statistics}}}.
\newblock {Springer}, {New York, NY}, 2012.
\newblock ISBN 978-1-4614-3654-6 978-1-4614-3655-3.
\newblock \doi{10.1007/978-1-4614-3655-3}.

\bibitem[James et~al.(2021)James, Witten, Hastie, and
  Tibshirani]{jamesIntroductionStatisticalLearning2021}
Gareth James, Daniela Witten, Trevor Hastie, and Robert Tibshirani.
\newblock \emph{An {{Introduction}} to {{Statistical Learning}}: With
  {{Applications}} in {{R}}}.
\newblock Springer {{Texts}} in {{Statistics}}. Springer US, 2021.
\newblock \doi{10.1007/978-1-0716-1418-1}.

\bibitem[Ramsay and Silverman(2005)]{ramsayFunctionalDataAnalysis2005}
J.~O. Ramsay and B.~W. Silverman.
\newblock \emph{Functional {{Data Analysis}}}.
\newblock Springer {{Series}} in {{Statistics}}. {Springer}, {New York, NY},
  2005.
\newblock ISBN 978-0-387-40080-8 978-0-387-22751-1.
\newblock \doi{10.1007/b98888}.

\bibitem[Ramsay et~al.(2023)Ramsay, Hooker, and
  Graves]{ramsayFdaFunctionalData2023}
James Ramsay, Giles Hooker, and Spencer Graves.
\newblock Fda: {{Functional Data Analysis}}, May 2023.

\bibitem[Song and Kim(2022)]{songSparseMultivariateFunctional2022}
Jun Song and Kyongwon Kim.
\newblock Sparse multivariate functional principal component analysis.
\newblock \emph{Stat}, 11\penalty0 (1):\penalty0 e435, 2022.
\newblock ISSN 2049-1573.
\newblock \doi{10.1002/sta4.435}.

\bibitem[Warmenhoven et~al.(2019)Warmenhoven, Cobley, Draper, Harrison,
  Bargary, and Smith]{warmenhovenBivariateFunctionalPrincipal2019}
John Warmenhoven, Stephen Cobley, Conny Draper, Andrew Harrison, Norma Bargary,
  and Richard Smith.
\newblock Bivariate functional principal components analysis: Considerations
  for use with multivariate movement signatures in sports biomechanics.
\newblock \emph{Sports Biomechanics}, 18\penalty0 (1):\penalty0 10--27, January
  2019.
\newblock ISSN 1476-3141.
\newblock \doi{10.1080/14763141.2017.1384050}.

\end{thebibliography}

\end{document}